\documentclass{aims}
\usepackage{amsmath}
  \usepackage{paralist}
  \usepackage{graphics} 
  \usepackage{epsfig} 
 \usepackage[colorlinks=true]{hyperref}
\hypersetup{urlcolor=blue, citecolor=red}

\def\currentvolume{X}
 \def\currentissue{X}
  \def\currentyear{200X}
   \def\currentmonth{XX}
    \def\ppages{X--XX}

\theoremstyle{definition}

\title[Laboratory Experiments on Wave Turbulence]
      {Laboratory Experiments on\\ Wave Turbulence}
\author[Eric Falcon]{}



 \email{eric.falcon@univ-paris-diderot.fr}

\begin{document}
\maketitle

\centerline{\scshape Eric Falcon }
\medskip
{\footnotesize
 \centerline{Laboratoire Mati\`ere et Syst\`emes Complexes (MSC)}\centerline{Universit\'e Paris Diderot, CNRS (UMR 7057)}
   \centerline{10 rue A. Domon \& L. Duquet, 75 013 Paris, France}
} 

\bigskip


\begin{abstract}
This review paper is devoted to a presentation of recent progress in wave turbulence. I first present the context and state of the art of this field of research both experimentally and theoretically. I then focus on the case of wave turbulence on the surface of a fluid, and I discuss the main results obtained by our group: caracterization of the gravity and capillary wave turbulence regimes, the first observation of intermittency in wave turbulence, the occurrence of strong fluctuations of injected power in the fluid, the observation of a pure capillary wave turbulence in low gravity environment and the observation of magnetic wave turbulence on the surface of a ferrofluid. Finally, open questions in wave turbulence are discussed.
\end{abstract}

\section{Context and state of the art}
\label{part1:intro}
When strong enough surface waves propagate in a medium, their interactions can generate waves of different wavelengths. This energy transfer through the different spatial scales can occur on a wide range of wavelengths. This stationary out-of-equilibrium state is called the wave turbulence: the energy of the system cascades through the scales from a scale where energy is injected up to a small scale where energy is dissipated. Wave turbulence thus concerns the study of dynamical and statistical properties of an ensemble of interacting nonlinear waves. The archetype of wave turbulence is the study of the random state of ocean surface waves generated by wind or current. 
But this is a very common phenomenon that is present in various situations:  surface waves on the sea \cite{Donelan85,Forristall81,Kahma81,Toba73}, internal waves in the ocean \cite{Lvov04}, Alfv\'en waves in solar winds \cite{Sagdeev79}, radar waves in ionosphere, spin waves in solids, Rossby waves in geophysics, ion waves \cite{Mizuno83} and Langmuir waves \cite{Huba80,Yu82} in plasmas, waves in nonlinear optics \cite{Kuznetsov91}, quantum waves in Bose condensates \cite{Berloff02,Conn05,Dyachenko96,Rica07} ... 

Wave turbulence is thus an interdisciplinary subject that involves different community: astrophysics, geophysics, mathematics, fluid mechanics and different fields of physics. Since the early developments of theoretical tools by applied mathematicians and physicists, the first ones to be interested on wave turbulence were oceanographers and meteorologists. Their motivations are numerous such as to develop climatic models, predict the sea state with more precision, or extract some energy from ocean waves as a source of alternative energy... At a more fundamental level, the goal is to understand the energy transfers between nonlinear interacting waves by means of generic laws whatever the particular medium in which these waves propagate.
 
Contrarily to a widespread belief, the analogy between wave turbulence and hydrodynamic turbulence within a fluid (3D turbulence) or within a film of fluid (2D turbulence) is quite limited. Although being also governed by the nonlinear effects and studied in a statistical and non-determinist way, wave turbulence is described by equations which strongly differs from Navier--Stokes ones of usual turbulence.
 
Wave turbulence theory, also known as weak turbulence (WT), is a statistical theory describing an ensemble of weakly nonlinear interacting waves. The first analytical studies began in the sixties \cite{Benney66,Benney67,Hasselmann62}. They were motivated by oceanographic measurements of Fourier spectra of the surface wave elevation as a function of the wavenumber. The aim was to understand how nonlinear interactions between the waves generate transfer of energy among different scales thus explaining the shapes of the observed Fourier spectra. In the early seventies, Zakharov and co-workers understood that, besides equilibrium spectra, nonlinearly interacting waves also involve spectra that correspond to the transfer of a finite energy flux from the large scales to small ones \cite{Zakharov67Grav,Zakharov67Cap,ZakharovLivre}. These stationary out-of-equilibrium solutions carrying fluxes of conserved quantities are the analogue of the Kolmogorov spectra of hydrodynamic turbulence. However, in the case of interacting waves, they can be computed analytically. To wit, let us consider a wave with a wavenumber $\vec{k}$ and a frequency $\omega_k$.  Its energy $E_{\vec{k}}$ can be read,
\begin{equation}
E_{\vec{k}} = n_{\vec{k}} \ \omega_k
\end{equation}
where $n_{\vec{k}}$ is called the ``wave action''.  WT then leads to the expression of the temporal evolution of the spectral density of the wave action by means of a kinetic equation such as
\begin{equation}
\frac{\partial n_{\vec{k}}}{\partial t} = C_{\vec{k}} - D_{\vec{k}} + I_{\vec{k}}.
\label{cinetique}
\end{equation}
where $C_{\vec{k}}$ is an interaction integral between waves ({\em e.g.} analogue of the collision integral of the Boltzmann formalism of the kinetic theory of gases), $D_{\vec{k}}$ a damping term representing the energy dissipation and $I_{\vec{k}}$ a forcing term describing the energy injection which generates the waves. 
When the nonlinearity in $C_{\vec{k}}$ is weak and the number of excited modes is large, a statistical description of wave turbulence makes sense. Indeed, by means of ensemble averages, the combination of weak nonlinearity and dispersion ensures that the long time asymptotics of the probability distribution of the wave field are universal (close to a Gaussian) for a broad band of initial distributions.  This means, that over sufficiently long times, the system relaxes towards Gaussian statistics insensitive to the initial probability distribution that is unkown {\it a priori}. This is called the asymptotic closure. The stationary solutions of the kinetic equations of weak turbulence can be then computed analytically at the equilibrium or in an out-of-equilibrium regime using perturbation methods of a small parameter quantifying the strength of the nonlinear term relative to the linear one. The stationary solutions are found scale invariant for isotropic and homogeneous systems by using a set of conformal transformations \cite{ZakharovLivre}. At the equilibrium ($D_{\vec{k}} = I_{\vec{k}} \equiv 0$), the classical solution is the so-called Rayleigh-Jeans spectrum corresponding to the equipartition of energy between modes. An out-of-equilibrium solution ($D_{\vec{k}} \neq 0$, $I_{\vec{k}}\neq 0$) is tractable when the scale of the energy injection (large scale) and the scale of the dissipation (small scale) are assumed widely separated to obtain an inertial regime independent of both the injection and the dissipation.
Eq.\ (\ref{cinetique}) then has conserved quantities (energy flux $\varepsilon\equiv \frac{\partial}{\partial t}\int n_{\vec{k}}\omega_k d\vec{k}$ or ``particle'' flux). For the finite-energy flux solution, the non zero first order expansion of the interaction integral allows to exactly works out the spectral density of the wave action $n_{\vec{k}}$ or the spectral density of the energy $E_{\vec{k}}$ such as\footnote{Note that, for isotropic systems, a frequency power law spectrum corresponds to the spatial one of Eq. (\ref{cascade}) by using the dispersion relation.} 
\begin{equation}
E_{\vec{k}} \sim \varepsilon^{1/(N-1)}k^{-\alpha}  {\rm \  \ with  \ \ } \alpha >1  {\rm \ ,}
\label{cascade}
\end{equation}
where the exponent of the energy flux depends on the $N$-wave interaction process which is itself fixed by both the wave dispersion relation and the dominant nonlinear interaction\footnote{For instance, the first non-zero resonant term for gravity waves is a $N=4$ wave interaction process, whereas for capillary waves $N=3$.}. This out-of-equilibrium solution is called the Kolmogorov-Zakharov spectrum (by analogy with the power-law spectrum obtained dimensionally for usual turbulence). Consequently, a direct ``energy cascade'' occurs: 
due to the interactions between nonlinear waves, the energy flux injected at large scale is transfers towards smaller scales. Similarly, for the finite-particle flux solution, the wave action $\mathcal{N}\equiv \int n_{\vec{k}}d\vec{k}$ is conserved, an invariant scale solution is found and an inverse cascade occurs (from small scales to large scales)\footnote{This is the analogue, in 2D turbulence, of the enstrophy conservation for the direct cascade, and the energy conservation for the indirect cascade.}. These spectra have been exactly computed in nearly all fields of physics involving wave dynamics \cite{ZakharovLivre}.  A lot of data obtained by remote sensing of the atmosphere or the ocean as well as satellite measurements in astrophysics, have thus been analyzed using the framework of wave turbulence.

The main assumptions for the derivation of the weak turbulence theory are:
\begin{itemize}
\item A weak nonlinearity is required due to the perturbation methods: the nonlinear term has to be much smaller than linear one (the fast linear time scale is separated from the slow nonlinear time scale). In practice, this condition corresponds to waves with enough amplitudes but not too high. 
\item The size of the system is assumed infinite. No boundary effect is taken into account theoretically.
\item The conserved quantity (energy flux) in wave turbulence is assumed constant with no fluctuation during the cascade through the scales. 
\item The hamiltonian formulation of weak turbulence assumes no dissipation at least in a transparency window in between the scale of injection (usually assumed at $\vec{k} =0$ or at the scale of the system), and the scale of dissipation at $\vec{k} \rightarrow \infty$.  In practice, this means that the forcing of waves has not to be at all scales.
\item The system is supposed isotropic and homogeneous spatially, and no coherent structures\footnote{For example, in hydrodynamics, vortices, cusps or wave breakings.} are involved.  
\item Waves have random phases. This conditions allows to derive a statistical equation for the ensemble averaged correlations of wave amplitudes.
\item The single assumption on the initial distribution of the wave field is that its cumulants have to decay at infinity.
\end{itemize}
At the moment, theoretical works try to model some finite size effects \cite{Kartashova98,Kartashova07,Kartashova08,Nazarenko06,Tanaka04,Zakharov05}. The role of coherent structures on the dynamics also begins to take into account theoretically \cite{Connaughton03}. During the last decade, theoretical or numerical works go beyond spectrum prediction. Transient behaviors during the evolution towards the stationary regime (or decaying wave turbulence) \cite{Kolmakov04,Kolmakov06}, condensation of classical waves (and its analogy with Bose condensation) \cite{Berloff02,Conn05,Dyachenko96,Korotkevitch08,Rica07} has been performed as well as the introduction of an empirical dissipation to model some strongly nonlinear processes (such as wave breakings) \cite{Zakharov07}.
 
In spite of the numerous analytical predictions for more than 40 years, wave turbulence is not so much studied experimentally. This is thus the opposite situation of the usual turbulence in which numerous experimental results exist, but very few analytical results can be carefully derived from master equations.

Oceanography and atmospheric sciences provide more and more {\em in situ} data nowadays given by weather balloons, surface buoys, radar and satellite measurements  \cite{Ochi05,Wiserevue07}. However, these large-scale systems depend on numerous parameters (namely for the ocean surface waves: wind, oceanic current, fetch...) leading to empirical determinations of the energy spectral density \cite{Ochi05}. Wind on the ocean surface also generates a wave forcing at various scales, and thus without scale separation as imposed by the theory. Laboratory experiments are much more relevant to accurately tune and control the system parameters. Surprisingly, such laboratory experiments are rare, but are fundamental in order to compare them with {\em in situ} measurements and theoretical predictions of weak turbulence. Only few laboratory experiments have been performed since 1996, mostly on wave turbulence on the surface of a fluid \cite{Brazhinikov02,Brazhinikov02bis,Henry00,Holt96,Lommer02,Wright96,Wright97}.  These works were focused on the measurement of the frequency power spectrum of capillary waves in roughly good agreement with the theoretical prediction of weak turbulence \cite{Zakharov67Cap}. Note that, some recent experiments of elastic wave turbulence on the surface of a thin plate \cite{Boudaoud08,Mordant08} show that the energy spectrum is in strong disagreement with the predictions of weak turbulence \cite{During06}.

In this paper, I will focus on the hydrodynamic case of wave turbulence. In \S \ \ref{hydro}, I will first review the previous results on this issue. Then, the main results of our experimental works on wave turbulence on the surface of a fluid will be described in \S \ \ref{resultats}. Finally, \S \ \ref{objectifs} will be devoted to the conclusion and a discussion on open questions in wave turbulence.  

\section{Hydrodynamic wave turbulence}
\label{hydro}
Waves on the surface of a fluid are in a gravity regime when their wavelength $\lambda$ is large with respect to the capillary length $\lambda_c \equiv 2\pi l_c=2\pi\sqrt{\gamma/\rho g}$ where $g$ is the acceleration of gravity, $h$ the fluid depth, $\rho$ the density and $\gamma$ the surface tension. Otherwise, they are in a capillary regime. The crossover between gravity and capillary regimes occurs for a wavelength close to the centimeter whatever the working fluid. The dispersion relation of linear inviscid surface waves reads $\omega^2(k)=\left[gk+(\gamma/\rho)k^3\right]\tanh{(kh)}$. In a deep fluid approximation, when $\lambda \ll 2\pi h$, one has
\begin{equation}
\omega^2=gk+\frac{\gamma}{\rho}k^3
\label{RD}
\end{equation}
By balancing both terms of the second hand of Eq.\ (\ref{RD}), one finds the critical wavenumber $k_c\equiv 1/l_c $ for the transition between both regimes. If waves are weakly nonlinear, the weak turbulence theory leads  to the expression of the power spectrum of surface wave amplitude such as
\begin{align}
& S^{grav}_{\eta}(\omega) \propto \varepsilon^{\frac{1}{3}}\ g\ \omega^{-4} & {\rm for\ gravity\ waves\ }  \label{WKTg} [97,103] \\ 
& S^{cap}_{\eta}(\omega) \propto \varepsilon^{\frac{1}{2}} \left(\frac{\gamma}{\rho}\right)^{\frac{1}{6}} \omega^{-\frac{17}{6}} & {\rm for\ capillary\ waves\ } [98] 
\label{WKT}
\end{align}

These relations can be also computed by dimensional analysis. In the gravity regime, parameters involved are the power spectrum of wave amplitude $S_\eta$, the gravity $g$, the energy flux $\varepsilon$ and the frequency $\omega$, of dimensions: $\left[S_{\eta}  \right] = L^2 T$ ; $\left[\varepsilon  \right]= L^3 T^{-3}$ ; $\left[g  \right]= L T^{-2}$ ; $\left[\omega  \right]= T^{-1}$. One can conclude by dimensional analysis only if the scaling law between $S^{grav}_{\eta}$ and $\varepsilon$ is assumed. As explained above in \S \ref{part1:intro}, this scaling is known as soon as the number $N$ of interacting waves is known [{\em i.e}, the order of the first non-zero term in the interaction integral $C_{\vec{k}}$ of Eq.\ (\ref{cinetique})]. Thus, for a $N$-wave resonant process, the power spectrum scales as $\varepsilon^{1/(N-1)}$. Gravity waves involving a 4-wave interaction process, one has $S^{grav}_{\eta} \propto \epsilon^{1/3}$.  One thus finds dimensionally the power spectrum of gravity wave as Eq.\ (\ref{WKTg}). Similar dimensional analysis can be performed for capillary waves assuming a 3-wave interaction process. Note that this dimensional analysis can be done whatever the wave dispersion relation of the system \cite{Connaughton03}.

Ocean surface waves are mainly generated by winds and currents at various scales. In the literature, {\it in situ} measurements of the exponent of the frequency-power law of the wave amplitude spectrum varies considerably, even if certain measurements roughly leads to a $\omega^{-4}$ scaling \cite{Donelan85,Forristall81,Kahma81,Toba73} thus suggesting a possible agreement with weak turbulence theory \cite{Zakharov67Grav,Zakharov82} and numerical simulations \cite{Dyachenko04,Onorato02,Pushkarev03} of gravity wave turbulence. However, the wave spectrum depends on numerous uncontrolled parameters such as duration of wind blowing, fetch length, stage of storm growth and decay, and existence of a swell, leading to fit the spectrum with too many parameters \cite{Ochi05}.

When gravity wave turbulence is not forced by wind, mechanisms of energy cascade in the inertial regime change, as well as the scaling law of the spectrum \cite{Kitaigorodskii83,Kuznetsov04,Onorato02}. Indeed, if $S_{\eta}$ does not depend on the energy flux, Phillips has shown dimensionally since 1958 that $S_{\eta}(\omega) \propto \varepsilon^{0}g^{2}\omega^{-5}$ \cite{Phillips58}. This situation corresponds to gravity waves of high amplitudes dissipating all the injected power (by wave breakings for instance). More recently, Kuznetsov has taken into account the presence of possible sharp wave-crests (cusps) on the surface that are assumed to propagate without deformation ({\it i.e.} using $\omega \propto k$ instead of $\omega=\sqrt{gk}$). If these sharp-crested structures occur along ridges (lines) then $S_{\eta}(\omega) \propto \omega^{-4}$ \cite{Kuznetsov04}. This exponent is similar to the one in Eq.\ (\ref{WKTg}) computed by weak turbulence theory (where no crested waves are involved). In the same way, when these wave slope divergences are assumed to be isolated peaks (points) propagating as $\omega=\sqrt{gk}$, the Phillips spectrum is found again \cite{Phillips58}. Recently, theoretical and numerical attempts try to take into account the quantization of wavenumber $k$ in finite size systems that can strongly affect energy transfers \cite{Kartashova98,Kartashova07,Kartashova08,Nazarenko06,Tanaka04,Zakharov05}. Notably, for gravity waves in a tank of characteristic size $L$, Nazarenko predicts that $S_{\eta}(\omega) \propto \varepsilon^{0}g^{5/2}L^{-1/2}\omega^{-6}$ \cite{Nazarenko06}. 

Experimentally, the capillary wave turbulence regime has been observed by means of optical methods \cite{Brazhinikov02,Brazhinikov02bis,Henry00,Holt96,Lommer02,Wright96,Wright97} and with a parametric forcing of the waves by shaking the whole container holding the fluid. It has been reported that the height of the surface displays a power-law frequency spectrum in $\omega^{-17/6}$ in agreement with weak turbulence theory  \cite{Zakharov67Cap} and simulations \cite{Pushkarev96}. However, peaks and its harmonics (related to the parametric forcing) are observed on the spectrum with maximal amplitudes decreasing as a frequency power-law \cite{Brazhinikov02,Brazhinikov02bis,Snouck08,Wright96,Wright97}. The frequency-spectrum exponent estimated in this way is thus not very accurate. A recent study has even found an exponent in disagreement with weak turbulence underlying the difficulty to reach a wave turbulence regime with a parametric forcing \cite{Snouck08}. Finally, note that new experiments of gravity wave turbulence in a large tank have been recently carried out \cite{Denissenko07,Savelsberg06,VanDeWater08}.

\begin{figure}[t]
\centerline{
\begin{tabular}{cc}
\includegraphics[height=4.5 cm]{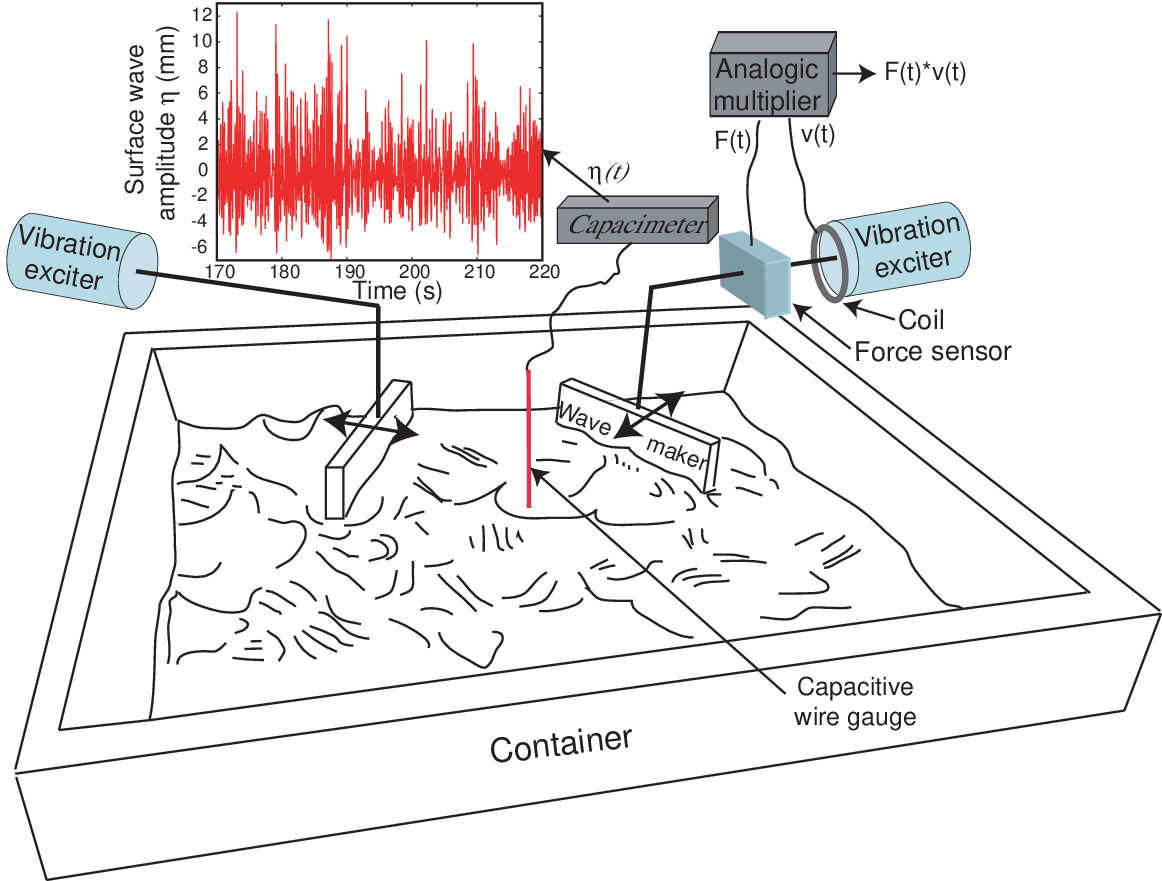}
&
\includegraphics[height=4 cm]{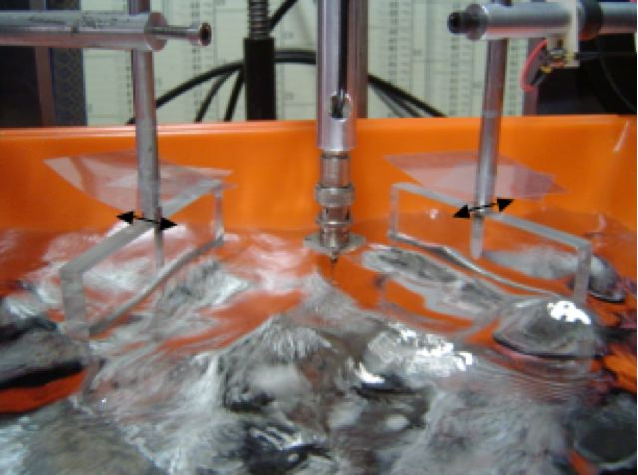}
\\
\end{tabular}}
\caption{Photo and schematic view of the experimental setup. Wave makers are on both sides of the capacitive sensor measuring the wave amplitude at a given location. Tank size: 20 cm $\times$  20 cm filled with water or mercury.}
\label{part1:fig01}
\end{figure}

\section{Overview of our results}
\label{resultats}
We have developed experiments to study wave turbulence on the surface of a fluid forced by wave makers \cite{Boyer08,Falcon07,Falcon07b,Falcon08,Falcon08sub1}. This investigation is of particular relevance since it allows to study on the same system two different regimes, those of gravity and capillary waves, that are characterized by different nonlinearities, 4-wave and 3-wave interactions respectively, leading to different poweer-law spectra. The crossover between the two regimes is also interesting, since it presents deviations from the weak turbulence theory\footnote{Close to this transition, the nonlinearity is not weak and the dynamics is dominated by fully nonlinear structures \cite{Connaughton03}.}. The role of the forcing parameters on the spectra had never been reported, nor the existence of a possible intermittency. The energy flux and its fluctuations had never been measured in wave turbulence, although it is the conserved quantity in wave turbulence. Another motivation is also to understand the statistical properties of energy flux necessary to maintain a dissipative system in an out-of-equilibrium stationary state. This is a general issue that is not restricted to the single case of wave turbulence.

\subsection{Spectrum and probability distribution of wave amplitudes \cite{Falcon07}}
\label{spectra}

As shown in Fig.\ \ref{part1:fig01}, surface waves are generated by two wave makers driven with a random excitation (both in frequency and amplitude) by means of electromagnetic vibration exciters. This random forcing is in a narrow low-frequency range selected by a low-pass filter (typically 0.1 -- 5 Hz). The square tank, 20 cm side, is filled either with water or mercury up to a height of 2 cm\footnote{Waves of wavelengths $\lambda \ll 2\pi h\simeq 12$ cm are thus in a deep-water regime.}. The wave amplitudes are measured at a given location by means of a capacitive method. This energy injected to the system by large wavelengths is transferred towards small structures due to nonlinearities. This process is characterized by measuring the frequency spectrum and the distributions of wave amplitude fluctuations. 

\begin{figure}[t]
\centerline{
\includegraphics[height=5 cm]{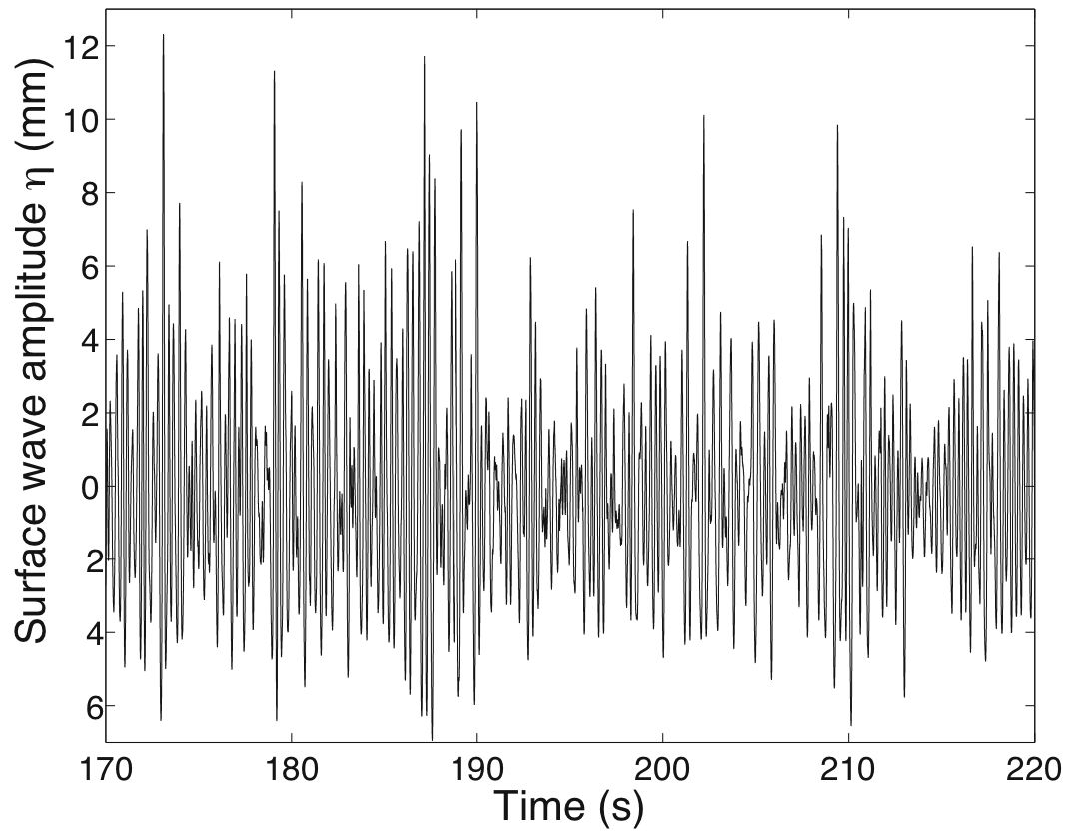}
}
\caption{Typical time recording of the surface waves, $\eta(t)$, during 50 s. $\langle \eta \rangle = 0$.}
\label{part1:fig02}
\end{figure}

A typical recording of the surface wave amplitude, $\eta(t)$, is displayed in Fig.\ \ref{part1:fig02} as a function of time. The signal is very erratic and the asymmetry observed is a signature of the steepness of the waves:  high crest waves are more probable than deep trough waves. By Fourier transform of such a signal, one can obtain the power spectrum density of the wave amplitudes  $S_{\eta}(\omega)\equiv \int \langle \eta(t+\tau)\eta(t)\rangle e^{-i\omega \tau}d\tau$. Waves being in nonlinear interactions, a wave turbulence regime is thus expected, that is the observation of a scale invariant spectrum.

\begin{figure}[t]
\centerline{
\includegraphics[height=5 cm]{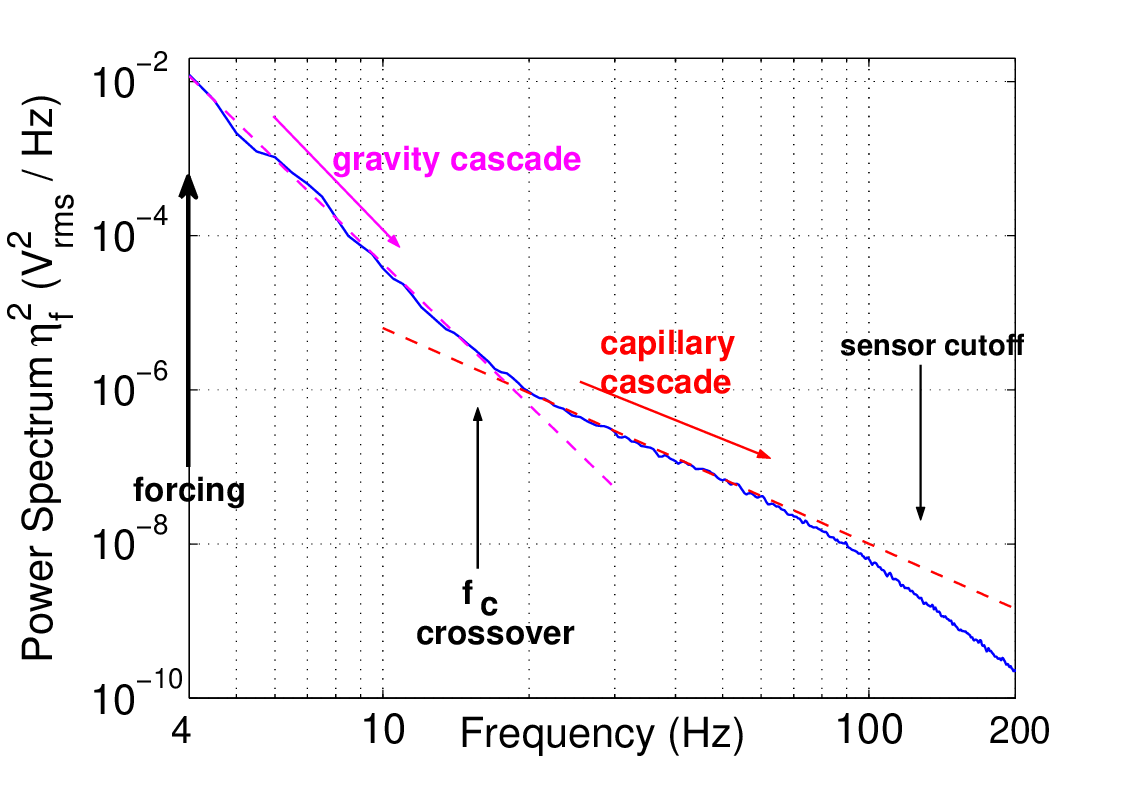}
}
\caption{Power spectrum of wave amplitude showing the regimes of gravity and capillary wave turbulence. The transition $\simeq 17$ Hz between both regimes corresponds to $\lambda_c = 2\pi l_c\simeq 1$ cm, where $l_c$ is the capillary length. Dashed lines have slopes $-6.1$ and $-3.2$. Random forcing with a narrow bandwidth $\leq 4$ Hz.}
\label{part1:fig03}
\end{figure}
\begin{figure}[ht]
\centerline{
\includegraphics[height=5 cm]{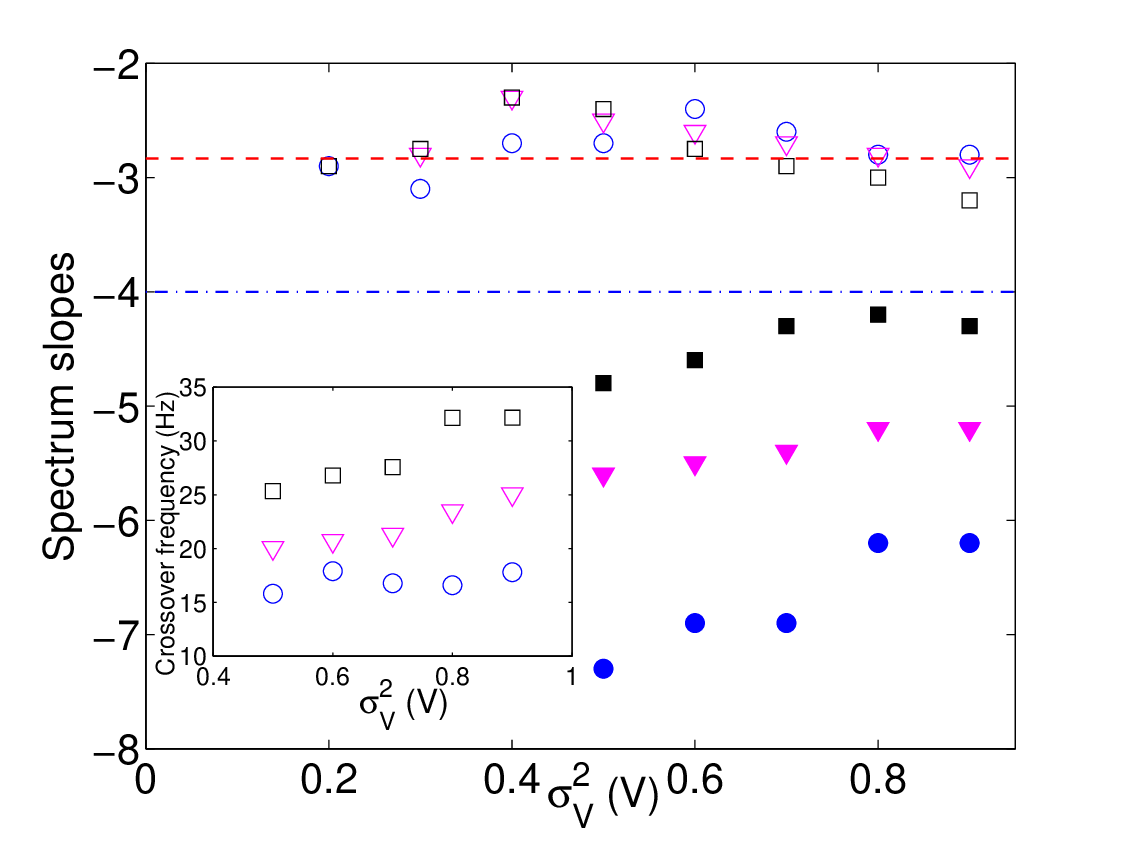}
}
\caption{Slopes of spectra of capillary (open symbols) and gravity (solid symbols) wave turbulence for different forcing amplitudes, $\sigma^2_V$, and bandwidths [($\circ$) 0 to 4 Hz, ($\bigtriangledown$) 0 to 5 Hz and ($\Box$) 0 to 6 Hz]. Dashed lines are weak turbulence predictions [see Eqs.\ (\ref{WKTg}) \& (\ref{WKT})]. {\sffamily Inset:} Crossover frequency between both regimes as a function of the forcing parameters.}
\label{part1:fig04}
\end{figure}
\begin{figure}[t]
\centerline{
\includegraphics[height=5 cm]{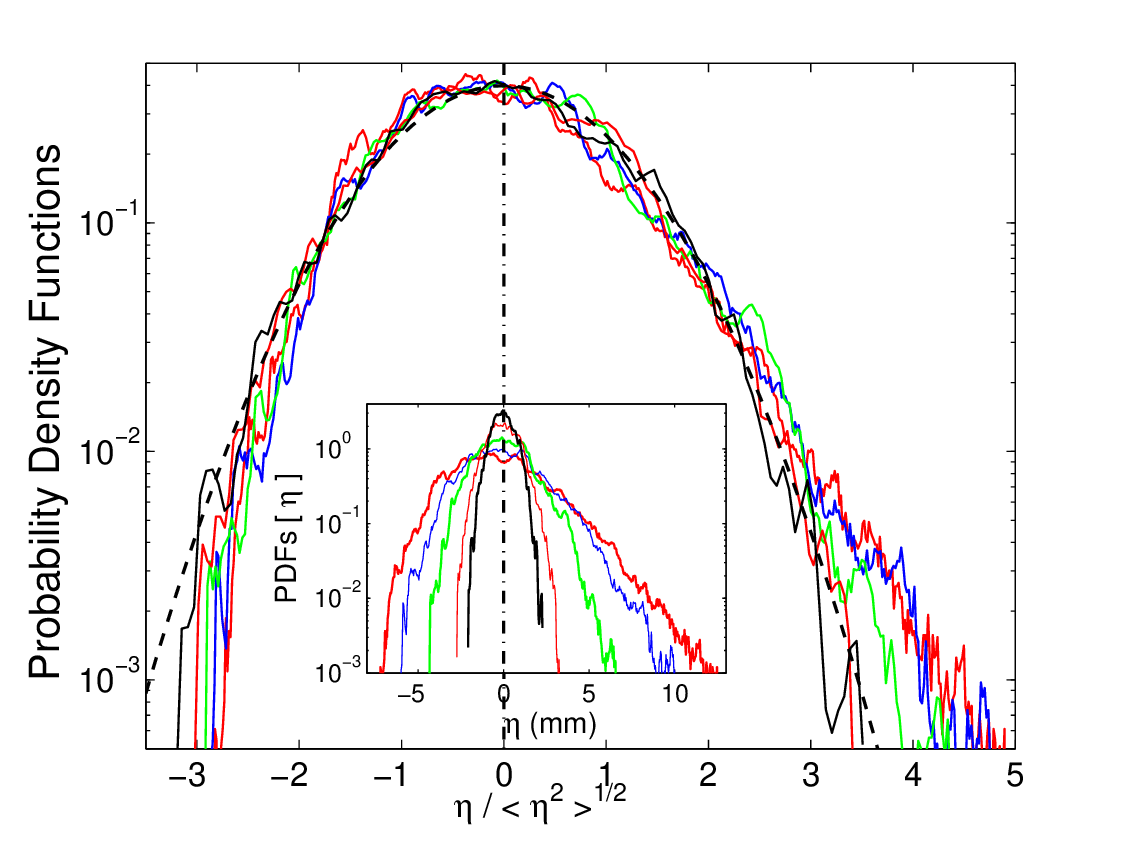}
}
\caption{{\sffamily Inset:} Probability density functions of the wave amplitude, $\eta(t)$, for different increasing amplitude of the forcing (from black to red curves). {\sffamily Main:} Same distributions rescaled by its standard deviation $\sqrt{\langle \eta^2 \rangle}$. Dashed lines: Gaussian fit with zero mean and unit standard deviation. }
\label{part1:fig05}
\end{figure}

As shown in Fig.\ \ref{part1:fig03}, we have observed, the crossover between the regimes of gravity (at large scales) and capillary (at small scales) wave turbulence  \cite{Falcon07}. Spectrum displays two different power laws corresponding to each regime. In the capillary regime, the frequency power-law exponent is in agreement with weak turbulence theory of Eq.\ (\ref{WKT}) in $\omega^{-17/6}$. Our estimation of the capillary-wave spectrum exponent  is much more accurate than in experiments using a parametric and monochromatic forcing \cite{Brazhinikov02,Brazhinikov02bis,Henry00,Holt96,Lommer02,Wright96,Wright97}. The transition between both regimes corresponds to a wavenumber of the order of the inverse of the capillary length $l_c$. For usual fluids, this corresponds to a critical frequency of $f_c=\sqrt{g/2l_c}/\pi \simeq 17$ Hz, that is a wavelength of the order of 1 cm. In the gravity regime, the spectrum exponent is found to be dependent on the forcing parameters (amplitude and bandwidth of the random forcing) as shown in Fig.\ \ref{part1:fig04}, and consequently in disagreement with the weak turbulence prediction of Eq.\ (\ref{WKTg}) in $\omega^{-4}$. The crossover frequency between both regimes also depends  on the forcing parameters (see inset of Fig.\ \ref{part1:fig04}). This dependence of the exponent of the gravity wave spectrum has been recently found again in a much larger tank \cite{Denissenko07}. Its origin is still an open problem. It could stem from finite size effects of the tank \cite{Zakharov05}. It has been also shown numerically that gravity wave spectrum is  very sensitive to the condensation of long waves background (or ``condensate'' following analogy with Bose-Einstein condensation in condensed matter physics) \cite{Berloff02,Conn05,Dyachenko96,Korotkevitch08,Rica07}. 

At low forcing amplitude, the probability density function (PDF) of the wave amplitudes is found to be Gaussian (see Fig.\ \ref{part1:fig05}).
At high enough forcing, the PDF becomes asymmetric thus showing the non-Gaussian feature of wave turbulence \cite{Falcon07}. This phenomenon is well known by oceanographers which measure crest-to-trough wave distributions which deviates from the usual Rayleigh distribution\footnote{A Gaussian process with negative and positive values has a Rayleigh distribution for its crest-to-trough amplitude.} \cite{Forristall00,Ochi05}. However, only few reports have mentioned this PDF asymmetry in laboratory experiments \cite{Onorato04}.

\subsection{Intermittency in wave turbulence \cite{Falcon07b}}
One of the most striking feature of turbulence is the occurrence of bursts of intense motion within more quiescent fluid flow. This generates an intermittent behavior \cite{Batchelor49,K62}. Indeed, a signal is called intermittent as soon as it displays more and more intense events when it is probed on a more and more short duration, leading thus to a strongly non Gaussian statistics at small time scales.
The origin of non-Gaussian statistics in 3D hydrodynamic turbulence has been ascribed to the formation of coherent structures (strong vortices) since the early work of  Batchelor and Townsend \cite{Batchelor49}. However, the physical mechanism of intermittency is still an open question that motivates a lot of studies in 3D turbulence \cite{Arneodo08,Chevillard07}. Intermittency has also been observed in a lot of problems involving transport by a turbulent flow [passive scalar, Burgers turbulence of strongly compressible fluid] \cite{Falkovitch01}, in magnetohydrodynamic turbulence in geophysics \cite{DeMichelis04} or in the solar wind \cite{Olga}. The possible observation of intermittency in wave turbulence that strongly differs from high Reynolds number hydrodynamic turbulence is of primary interest. It could indeed motivate explanations of intermittency different than the ones considering the dynamics of the Navier-Stokes equation.

\begin{figure}[t]
\centerline{
\includegraphics[height=5 cm]{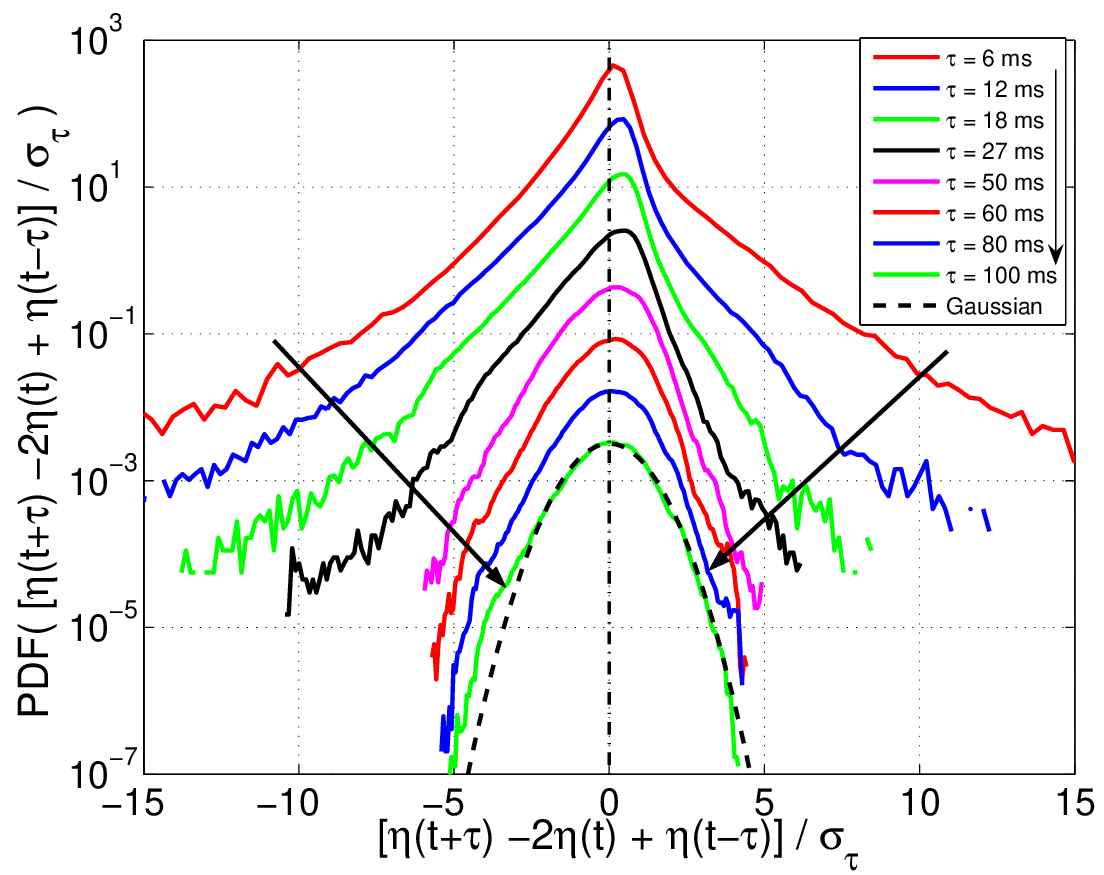}
}
\caption{Probability density functions of the local slopes of wave amplitudes $[\eta(t+\tau) -2\eta(t) + \eta(t-\tau)] / \sigma_{\tau}$ for different time lags $6 \leq \tau \leq 100$ ms (from top to bottom). Gaussian fit with zero mean and unit standard deviation (dashed line). Correlation time: $\tau_c\simeq 63$ ms. Each curve has been shifted for clarity.}
\label{part1:fig06}
\end{figure}
\begin{figure}[t]
\centerline{
\includegraphics[height=5 cm]{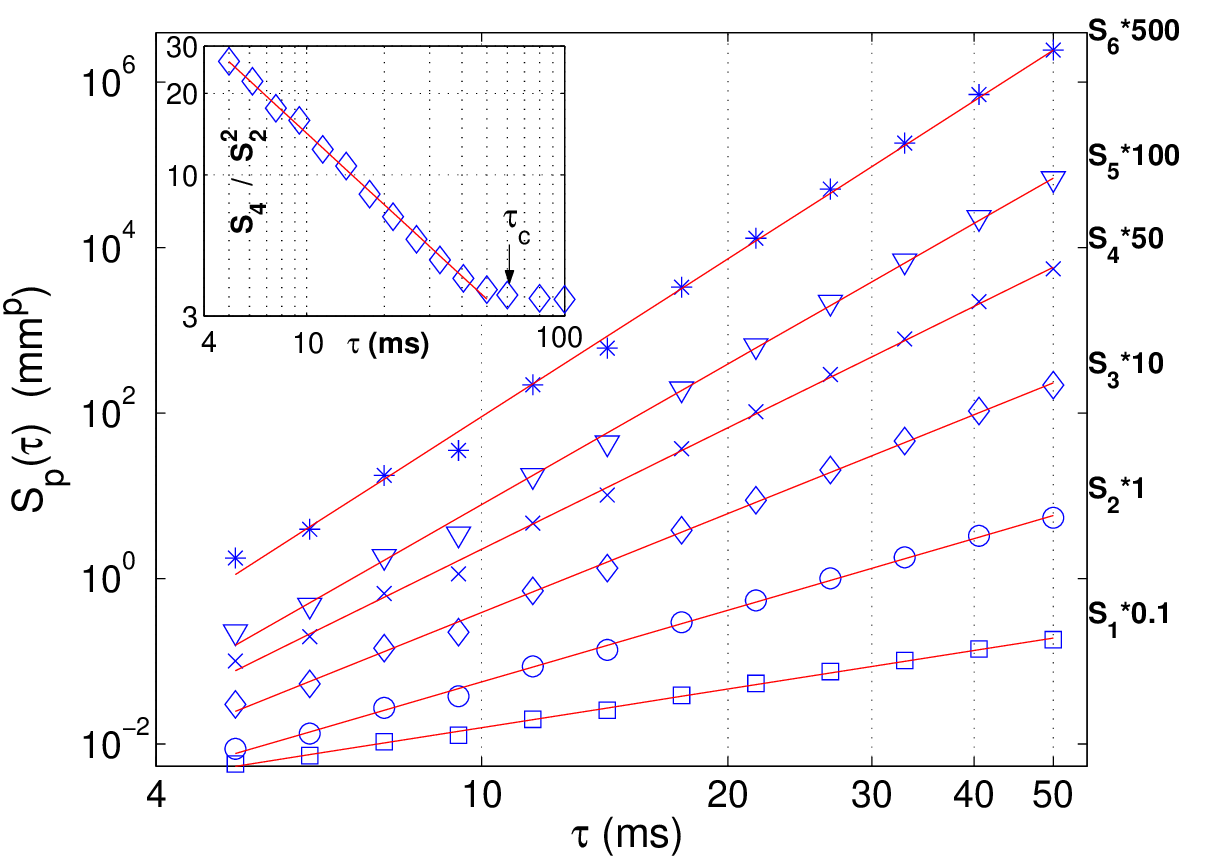}
}
\caption{Structure functions of order $p$ of the local slopes of wave amplitudes, ${\mathcal S}_p(\tau)$, as a function of time lag $\tau$, for $1\leq p \leq 6$. ($-$): Power-law fits, ${\mathcal S}_p(\tau) \sim \tau^{\xi_p}$, where the slopes $\xi_p$ depend on the order $p$ (see Fig.\ \ref{part1:fig08}). Curves have been shifted for clarity. {\sffamily Inset:} Flatness ${\mathcal S}_4/{\mathcal S}_2^2$ as a function of $\tau$ fitted by a power law of slope $-0.88$ ($-$). Correlation time: $\tau_c\simeq 63$ ms.}
\label{part1:fig07}
\end{figure}

\begin{figure}[t]
\centerline{
\includegraphics[height=5 cm]{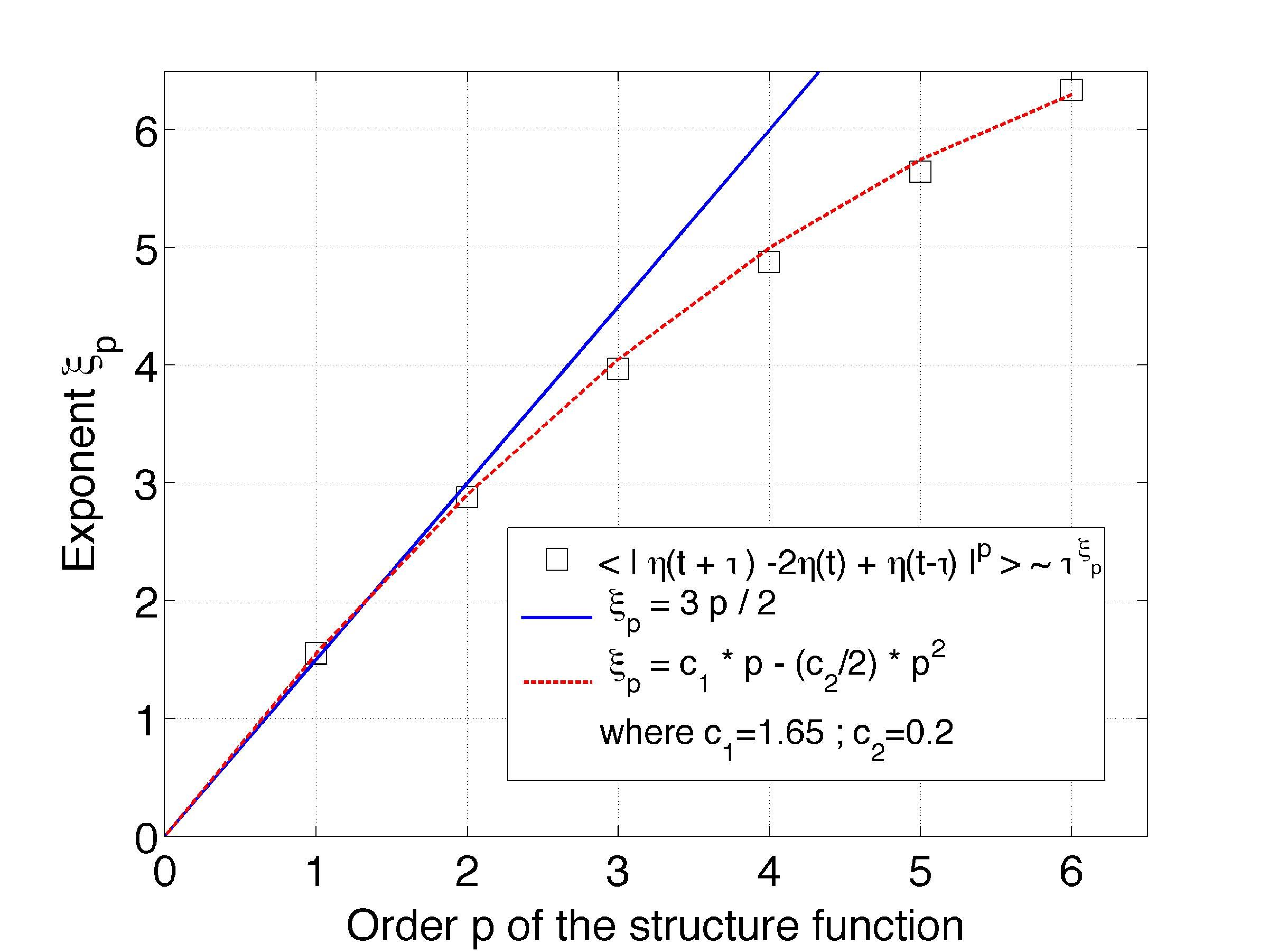}
}
\caption{Exponents $\xi_p$  of the structure functions as a function of the order $p$. $\xi_p$ is computed from the slopes of Fig.\ \ref{part1:fig07}, and is fitted by ($--$) $\xi_p=c_1p-\frac{c_2}{2}p^2$ with $c_1=1.65$ and $c_2=0.2$. Solid line: dimensional analysis $\xi_p=3p/2$.}
\label{part1:fig08}
\end{figure}

We have reported the first observation of intermittency in wave turbulence \cite{Falcon07b}. By measuring the temporal fluctuations of the surface wave amplitude, $\eta(t)$, at a given location (cf. Fig.\ \ref{part1:fig01}), one can compute the local slope increments of the gravity surface waves at a certain time scale $\tau$: ${\Delta}\eta(\tau)\equiv {\eta}(t+\tau)-2{\eta}(t)+{\eta}(t-\tau)$. As shown in Fig.\ \ref{part1:fig06}, starting from a roughly Gaussian statistical distribution of these increments at large scale $\tau$, the PDFs undergoes a continuous deformation towards strongly non-Gaussian shape at smaller scales. This shape deformation of the distributions across the scales is a direct signature of intermittency. Figure\ \ref{part1:fig07} 
shows the structure functions of these increments, defined as ${\mathcal S}_p(\tau) \equiv  \langle |\Delta\eta(\tau)|^p \rangle$, which are found to be power laws of the time scale $\tau$ on more than one decade, such as ${\mathcal S}_p(\tau) \sim \tau^{\xi_p}$ where $\xi_p$ 
is an increasing function of the order $p$. The evolution of the exponents $\xi_p$ with $p$ is found to be a nonlinear function of $p$ as shown in Fig.\ \ref{part1:fig08}. The nonlinearity of $\xi_p$ is another direct signature of intermittency. The slope at the origin of the curve $\xi_p$ vs. $p$ (the non-intermittent part) is close to $3p/2$, this value being easily derived by dimensional analysis\footnote{One has indeed $S_p(\tau)\sim \varepsilon^{p/6}g^{p/2}\tau^{3p/2}$.}. However, there exists any model that can describe the intermittent  nonlinear part of $\xi_p$ at large $p$.

Recently, it has been proposed that theoretical corrections should be taken into account in weak turbulence to describe a possible intermittency that may be connected to singularities or coherent structures \cite{Choi05,Connaughton03} such as wave breaking  \cite{Yokoyama04} or whitecaps \cite{Connaughton03} on the fluid surface. However, intermittency in wave turbulence is often related to non-Gaussian statistics of low wave number Fourier amplitudes \cite{Choi05}, and thus it is not obviously related to small-scale intermittency of usual hydrodynamic turbulence. Here, we have shown the intermittent feature of the local slope of turbulent gravity waves in a same way as intermittency in usual turbulence. A challenge is to understand the physical origin of intermittency in wave turbulence. 

\subsection{Fluctuations of the energy flux \cite{Falcon08,Falcon08sub2}}
\begin{figure}[ht]
\centerline{
\includegraphics[height=5 cm]{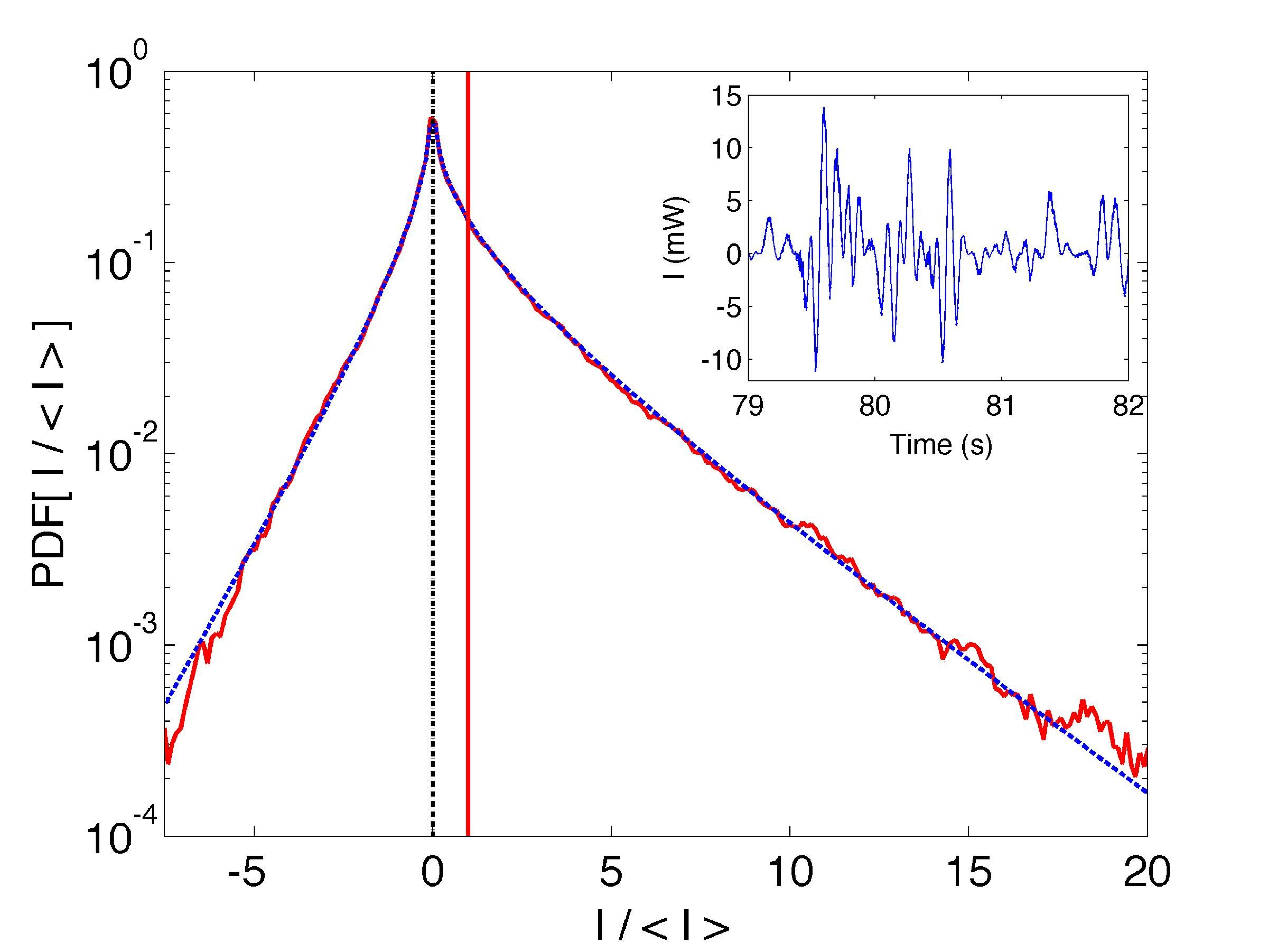}
}
\caption{{\sffamily Inset:} Time recording of injected power, $I(t)$, driven the surface waves. $\langle I \rangle$ = 2 mW. {\sffamily Main:} Probability density function of injected power showing two exponential tails. Its asymmetry is related to its mean value $\langle I \rangle$ (vertical solid line). Note that fluctuations are much more larger the mean value. Blue dashed line: prediction of our model with any adjustable parameter.}
\label{part1:fig11}
\end{figure}

\begin{figure}[t]
\centerline{
\includegraphics[height=5 cm]{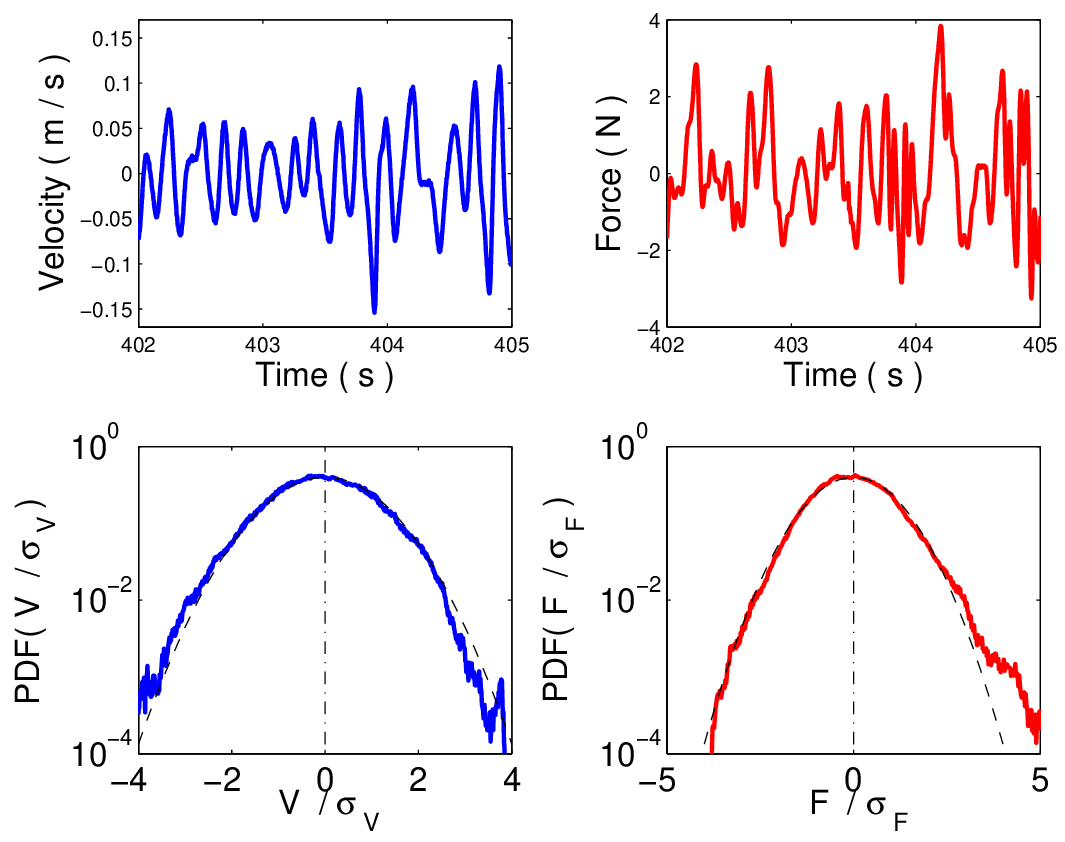}
}
\caption{Time recordings of the velocity $V(t)$ of the wave maker (left) and the force applied $F(t)$ to the wave maker by the vibration exciter (right). Both probability distributions are quasi Gaussian (see dashed lines) with zero mean value. $\sigma_V$ and $\sigma_F$ are the standard deviations.}
\label{part1:fig12}
\end{figure}

The key parameter of wave turbulence is the energy flux injected in the system. It is assumed to be conserved during the cascade across the scales. For the first time in wave turbulence, we have measured the instantaneous injected power in the fluid $I(t) \equiv F\times V$, where $F(t)$ is the force applied by the wave maker on the fluid, and $V(t)$ the wave maker velocity (see Fig.\ \ref{part1:fig01}). The typical temporal evolution of this energy flux is shown in the inset of Fig. \ref{part1:fig11}. It displays fluctuations much stronger than its mean value $\langle I \rangle$ (see Fig.\ \ref{part1:fig11}).  These fluctuations of energy flux have not been taken into account in the theoretical models that have been developed up to date. To wit, the knowledge of their evolution during the whole cascade would be necessary. Here, one studies the statistical properties of the energy flux fluctuations measured on the wave maker (integral scale). Numerous events of negative energy flux ($I(t)<0$) are observed on Fig.\ \ref{part1:fig11} with a fairly high probability. These are events for which the wave field gives back energy to the wave maker. Understanding this phenomenon is necessary to develop marine renewable energy such as ocean wave energy. It is thus of first interest to model this empirical statistical distribution of the energy flux. To wit, the wave maker is described by a Langevin-type model with a pink noise (Ornstein-Uhlenbeck process) that mimics the experimental narrow-band random forcing \cite{Falcon08}.  Two linear coupled Langevin-type equation for $V$ and $F$ are then obtained; these two variables being Gaussian (due to the forcing) as observed experimentally (see Fig.\ \ref{part1:fig12}). The computation of the distribution of the product of these two correlated Gaussian distributions thus gives the analytical expression of the energy flux distribution with no adjustable parameter \cite{Aumaitre08,Falcon08,Falcon08sub2}. Figure\ \ref{part1:fig11} shows a very good agreement between our model and the experiment. The asymmetry of the probability distribution is driven by the mean energy flux $\langle I \rangle$, that is equal, in a stationary state, to the mean dissipated power. This model is not restricted to wave turbulence, but is very generic since it also describes the energy flux distribution in other dissipative out-of-equilibrium systems such as turbulent convection, granular gases, bouncing ball with a random vibration \cite{Aumaitre08}, and electronic circuit with a random voltage (see below and Ref. \cite{Falcon08sub2}). 


We have also emphasized the bias that can result from the system inertia when one tries a direct measurement of the fluctuations of injected power. When the wave maker inertia is not negligible\footnote{For our experiments, this is the case with water but not with mercury.}, the measurement of the instantaneous force has to be corrected by the inertia before performing the product with the velocity signal in order to deduce the power. Otherwise, it may lead to wrong estimations of the fluctuations of injected power, the mean value being not obviously affected. There exists only a few previous direct measurements of injected power in 3D turbulent flows, and this type of inertial bias has never been taken into account \cite{vkbias,vkbias2,vkbias3}.

\begin{figure}[t]
\centerline{
\includegraphics[height=5 cm]{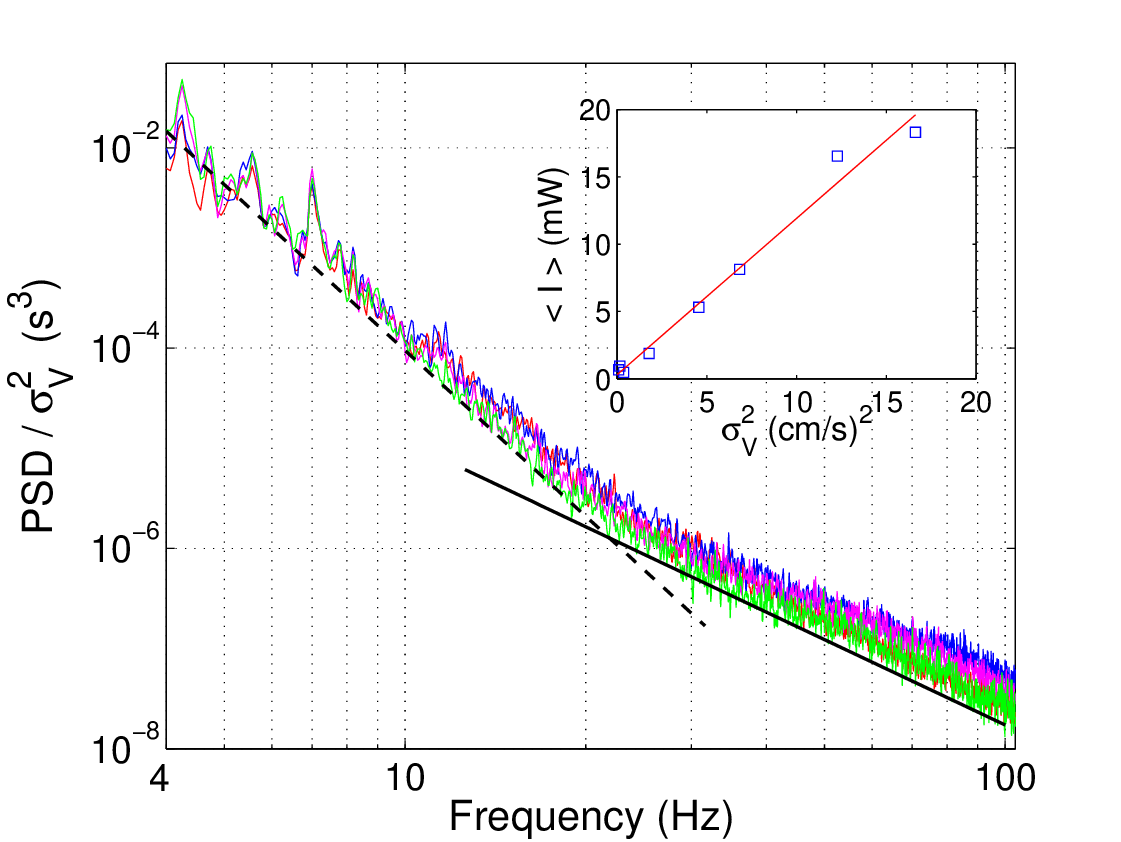}
}
\caption{Spectra of the surface wave amplitude divided by the variance $\sigma^2_V$ of the wave maker velocity for different forcing amplitudes: $\sigma_V = 2.1, 2.6, 3.5$ et $4.1$ cm/s. The dashed line has slope $-5.5$, whereas the solid line has slope $-17/6$. {\sffamily Inset :} The mean injected power $\langle I \rangle$ is found proportional to $\sigma_V^2$.}
\label{part1:fig13}
\end{figure}

The mean injected power $\langle I \rangle$ is found to be  proportional to the fluid density\footnote{This has been verified for only two different fluids (water and mercury) but with a density ratio close to 14.} $\rho$ and to $\sigma^2_V$, the rms value squared of the wave-maker velocity fluctuations. The rms value of the injected power fluctuations also scales as $\sigma_I \sim \langle I \rangle \sim \rho \sigma^2_V$. The energy transfer occurs via 4-wave interactions in the gravity regime and via 3-wave interactions in the capillary regime. Thus, if these two regimes are independent as in Eqs.\ (\ref{WKTg}) and (\ref{WKT}), the dependence of the spectrum amplitude $S_{\eta}$ on the mean energy flux $\langle I \rangle$ should display different scaling laws. Our measurements show that a single scaling law, $S_{\eta} \sim \langle I \rangle$, leads to the collapse on a single master curve of our experimental spectra for different values of the forcing (see Fig.\ \ref{part1:fig13}). This strong disagreement with the weak turbulence theory [cf. Eqs.\ (\ref{WKTg}) \& (\ref{WKT})] thus shows that interactions between these regimes are very important although they are not taken into account theoretically.

The experimental distribution of the injected power averaged on a time lag $\tau$, $P(I_{\tau})$, has also been studied \cite{Falcon08} in the framework of the Fluctuation Theorem of the out-of-equilibrium statistical physics (also called the Gallavotti-Cohen relationship) \cite{FT,FT2}. This theorem states that  $\frac{P[+I_{\tau}/\langle I \rangle]}{P[-I_{\tau}/\langle I \rangle]} \sim e^{-\beta(\tau) I_{\tau}/\langle I \rangle}$ where $P[x]$ is the probability to have the value $x$ and $\beta(\tau)$ is a constant related to the energy of the system. The theorem hypotheses (such as the temporal reversibility) are generally not fulfilled for a real dissipative system.  The apparent verification of this theorem in numerous previous experiments \cite{Ciliberto98,Ciliberto04,Feitosa04,Shang03,Shang05} 
is due to their small range of explored fluctuation amplitudes $I_{\tau}/\langle I \rangle$ at long averaging time $\tau$ \cite{Aumaitre99bis,Aumaitre99,Aumaitre04,Gilbert04,Puglisi05,Visco05}. Our wave turbulence experiment allows us to reach large enough fluctuations of injected power, and we have shown that they do not verify the theorem \cite{Falcon08}, but are found in good agreement with the analytical calculation performed with a Langevin-type equation forced with a Gaussian white noise \cite{Farago}.

\begin{figure}[ht]
\centerline{
\includegraphics[height=5 cm]{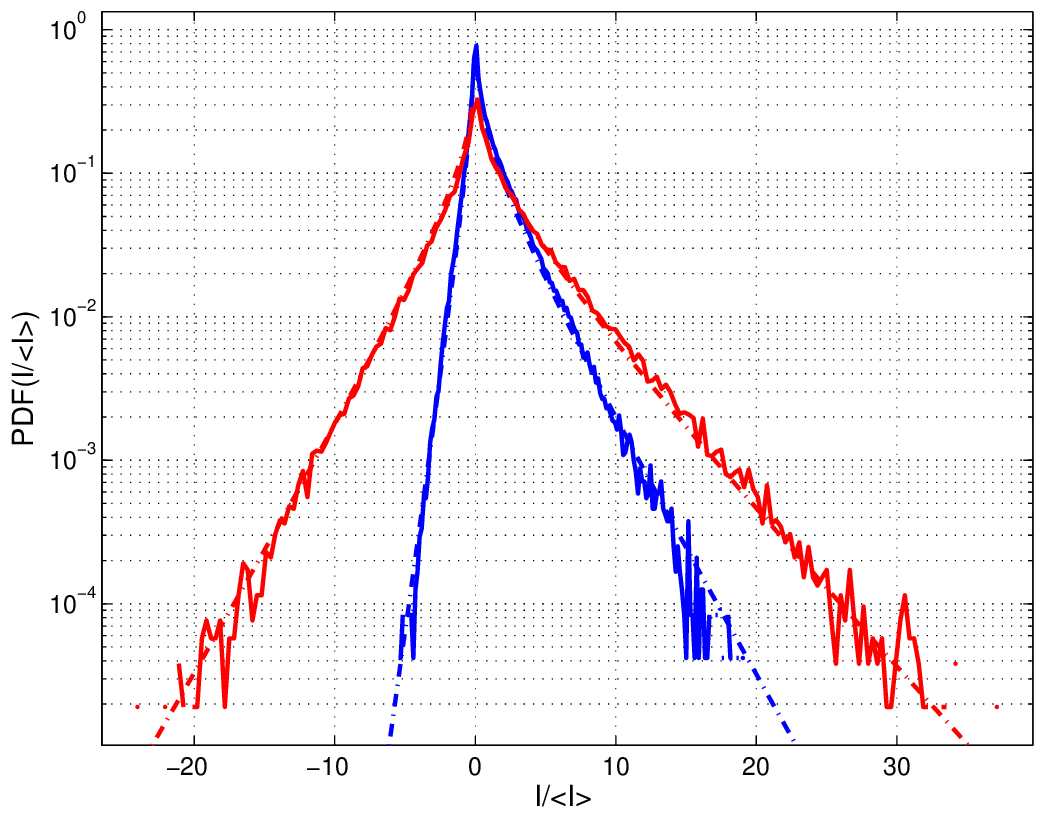}
}
\caption{Fluctuations of injected power $I(t)$ in an electronic RC circuit driven with a stochastic voltage for two different values of the damping rate $(RC)^{-1}= 2000$ and 200 Hz (resp. blue and red curves). Comparison between experiment ($-$) and theoretical prediction of the Langevin-type model ($--$). As in wave turbulence, the more dissipative the system is, the more asymmetric the PDF of $I(t)$ is.}
\label{part1:fig14}
\end{figure}

We have also studied experimentally the energy flux fluctuations in an electronic RC circuit driven with a stochastic voltage \cite{Falcon08sub2}. This out-of-equilibrium model system allows us to easily control the system parameters (notably, its damping rate $1/(RC)$). We show that the large fluctuations reached do not verify  the Fluctuation Theorem even for long averaging time $\tau$. As shown in Fig.\ \ref{part1:fig14}, the probability distribution of the instantaneous energy flux is qualitatively similar to the one of wave turbulence of Fig.\ \ref{part1:fig11}, and is indeed well described by our above model (see dashed lines in Fig.\ \ref{part1:fig14}). This electronic circuit is thus one of the simplest system to understand certain properties of the energy flux fluctuations shared by other dissipative out-of-equilibrium systems, such as in granular gases \cite{Feitosa04}, convection \cite{Shang03,Shang05}, or wave turbulence \cite{Falcon08}. 

\subsection{Wave turbulence in low-gravity environment \cite{Falcon08sub1}}
\begin{figure}[t]
\centerline{
\includegraphics[height=5 cm]{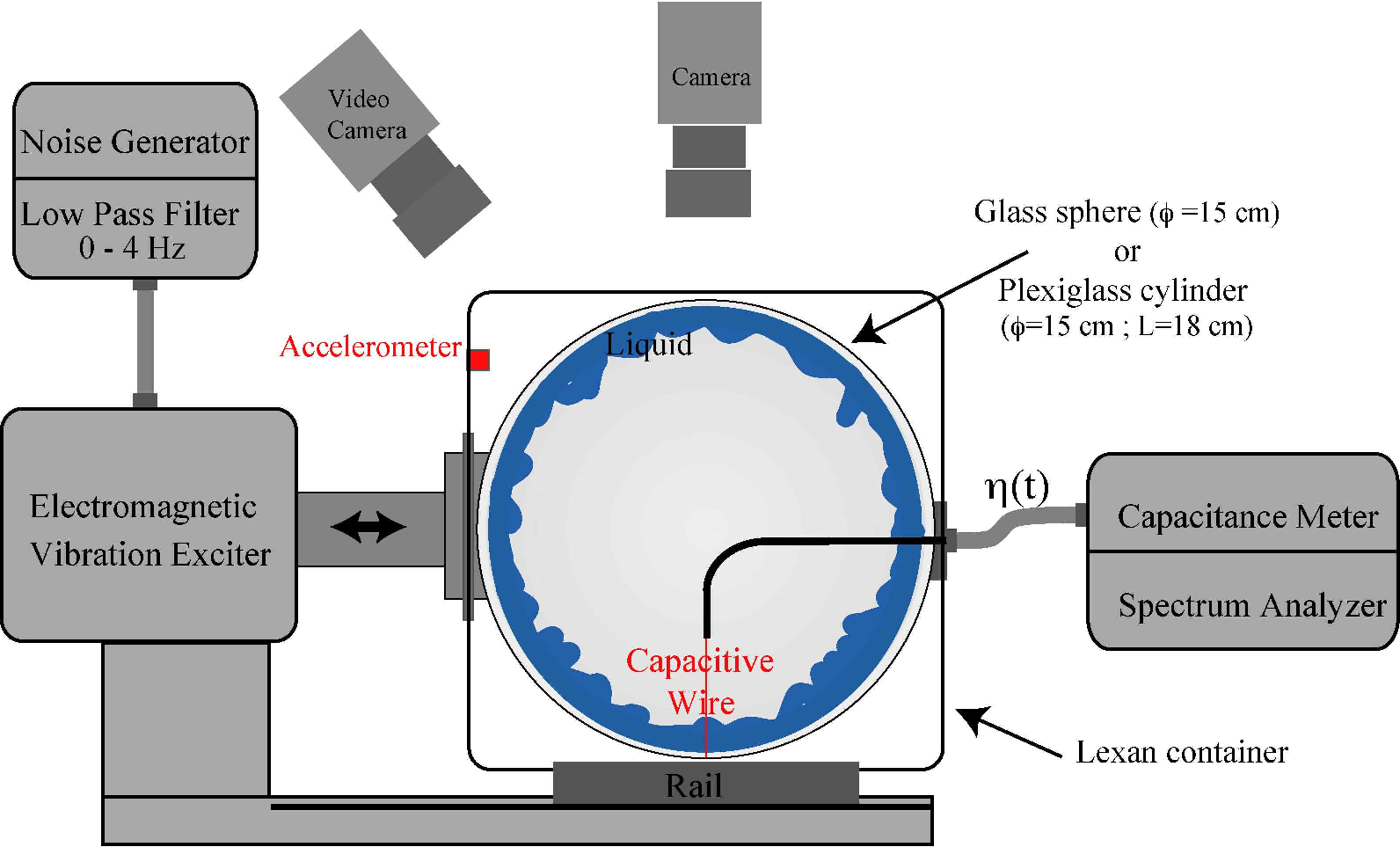}
}
\caption{Experimental setup for the wave turbulence in low gravity. In low-gravity periods, the fluid covers all the internal surface of the tank which is submitted to vibrations.}
\label{part1:fig15}
\end{figure}

Experiments of wave turbulence in low gravity are relevant to study pure capillary waves on a large range of wavelengths without interfering with the gravity waves. During ground experiments, the upper limit is the capillary length, and its lower limit is related to viscous dissipation. Our experiment in weightlessness has been carried out during CNES or ESA campaign of parabolic flights on board of an specially modified Airbus A300. A campaign consists of 3 flights, each flight consists of a series of 30 parabolic trajectories that result in low-gravity periods, each of 22 s. Such a low-gravity environment (about $\pm 5\times 10^{-2}$g) allows us to also study the behavior of a spherical fluid layer on which waves can be propagates with no boundary condition (see the experimental setup in  Fig.\ \ref{part1:fig15}).

\begin{figure}[t]
\centerline{
\includegraphics[height=5 cm]{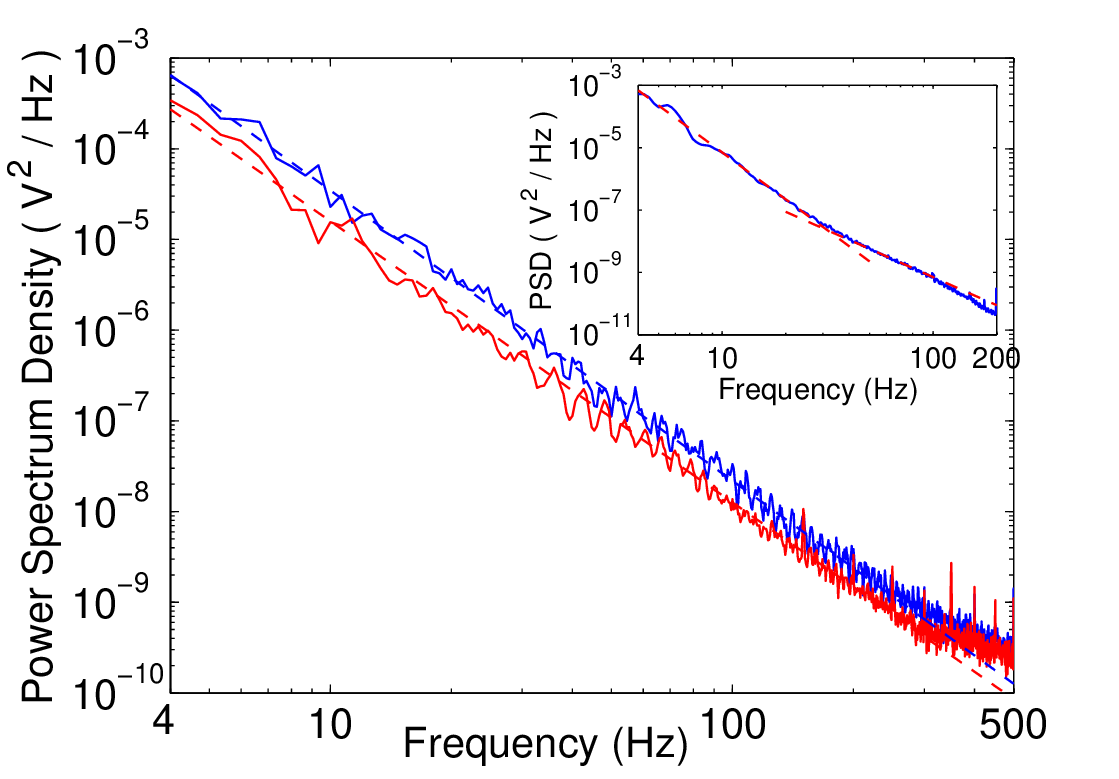}
}
\caption{Power spectra of surface wave amplitudes {\sffamily \em in low gravity} for two different forcings: random forcing from 0 to 6 Hz (lower), or sinusoidal forcing at 3 Hz (upper). Dashed lines have slopes of -3.1 (lower) et -3.2 (upper). {\sffamily Inset :} Same {\sffamily \em \ with gravity}. Two power laws appear corresponding to the gravity (slope -5) and capillary (slope -3) regimes.}
\label{part1:fig16}
\end{figure}

In such low-gravity conditions, we have observed the capillary wave turbulence on the surface of a fluid layer that covers all the internal surface of a spherical container which is submitted to random forcing \cite{Falcon08sub1}. Figure\ \ref{part1:fig16} displays the capillary wave spectrum over a large range of wavelengths, notably in wavelengths where gravity waves were present during ground experiments. This spectrum over two decades in frequency (corresponding to wavelength from $mm$ to few $cm$) is found in roughly good agreement with the frequency-exponent prediction of weak turbulence theory of the capillary regime in $-17/6 \simeq -2.8$ [see Eq.\ (\ref{WKT})]. As shown in Fig.\ \ref{part1:fig16}, this exponent is also found to be independent of the large-scale forcing parameters (low-frequency sinusoidal or random forcing) as expected within the inertial range.

\begin{figure}[t]
\centerline{
\includegraphics[height=4 cm]{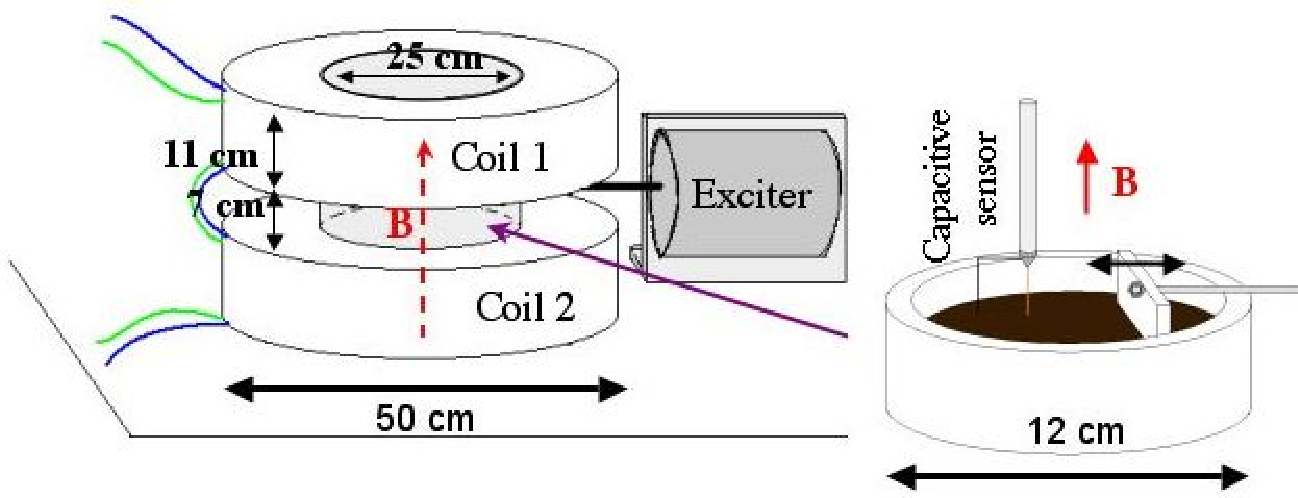}
}
\caption{Experimental setup for the study of wave turbulence on the surface of a ferrofluid in a normal magnetic field generated by two coaxial coils. Surface waves on the ferrofluid are generated by the horizontal motion of a plunging wave maker driven by an vibration exciter. The amplitude of surface waves at a given location is measured by a capacitive wire gauge.}
\label{part1:fig18}
\end{figure}

\subsection{Wave turbulence on the surface of a ferrofluid \cite{Boyer08}}
A ferrofluid is a colloidal suspension of nanometric ferromagnetic particles diluted in a liquid displaying thus both liquid and strongly magnetic feature. In contrast with usual liquids, the dispersion relation of surface waves on a ferrofluid is nonmonotonic and displays a minimum (depending on the amplitude of the applied magnetic field $B$ \cite{Broaweys99,Broaweys00,Embs,Mahr96,Muller99}) for which an instability occurs. Above a critical field $B_c$, a static hexagonal pattern of peaks occurs on the ferrofluid surface, the so-called Rosensweig instability \cite{Cowley67}. 
The amplitude of the magnetic field, $B$, can thus easily tune the dispersion relation of the surface wave which reads \cite{Cebers,Rosen} 
\begin{equation}
\omega^2=gk-\frac{f[\chi]}{\rho\mu_0}B^2k^2+\frac{\gamma}{\rho}k^3
\label{rdtheo}
\end{equation}
where $\mu_0=4\pi \times 10^{-7}$ H/m is the magnetic permeability of the vacuum, and $f[\chi]$ a known function of the magnetic susceptibility of the ferrofluid\footnote{Indeed, $\chi($H$)$ depends on the applied magnetic field, $H$, through Langevin's classical law, and thus on the magnetic induction $B$, since $B=\mu_0(1+\chi)H$. The magnetic induction, $B$, will be called hereafter the magnetic field for simplicity.}. One can thus easily tune, with just one single control parameter $B$, the dispersion relation of surface wave from a dispersive one ($k$ or $k^3$ term) to a nondispersive one ($k^2$ term). Besides, some theoretical questions are open notably about the possible existence of power-law solution for the energy wave spectrum for two-dimensional nondispersive systems \cite{Connaughton03,Lvov97,Newell71,Zakharov70}. Wave turbulence on the surface of a magnetic fluid should thus display striking properties depending on whether the system is below, close or above the onset of the Rosensweig instability.

\begin{figure}[t]
\centerline{
\includegraphics[height=5 cm]{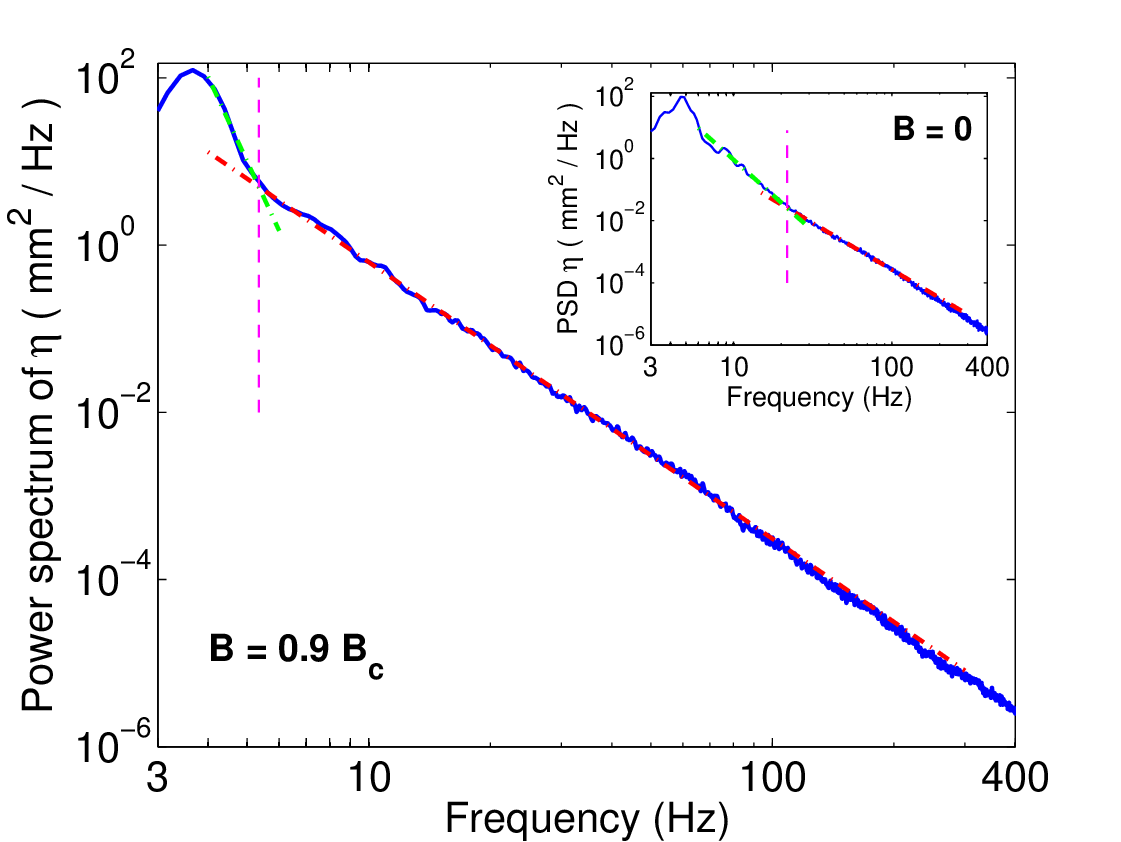}
}
\caption{Power spectrum of wave amplitude, $\eta(t)$, on the ferrofluid surface for two applied magnetic field. {\sffamily Inset:} $B=0$ : Gravity and capillary wave turbulence regimes. Dashed lines have slopes -4.6 and -2.9. Vertical dashed line: crossover frequency $f_{gc}$ Hz. {\sffamily Main:} $B=0.9B_c$ : Magnetocapillary wave turbulence. Dashed line has slope -3.3. Crossover is decreased towards $f_{gc}\simeq 5$ Hz. }
\label{part1:fig19}
\end{figure}

We have studied the wave turbulence on the surface of a ferrofluid submitted to a normal magnetic field (see Fig.\ \ref{part1:fig18}) \cite{Boyer08}. We have shown that magnetic surface waves arise only above a critical field, but below the onset $B_c$ of the Rosensweig instability. As shown in Fig.\ \ref{part1:fig19}, the power spectrum of the magnetic wave amplitudes displays a single power law over the whole range of accessible frequencies instead of two regimes (gravity and capillary) observed with no applied magnetic field. On can compute by dimensional analysis the power spectrum of magnetic wave turbulence as
\begin{equation}
S^{mag}_{\eta}(\omega) \sim \varepsilon^{\alpha}\left(\frac{B^2}{\rho\mu_0}\right)^{\frac{2-3\alpha}{2}}\omega^{-3} {\rm \ \ , }
\label{spectremag}
\end{equation}
This predicted frequency exponent $\omega^{-3}$ is too close to the one predicted for the capillary regime in $\omega^{-17/6}$ to be distinguished experimentally with our experimental accuracy. This could explain that only one single slope is observed  in Fig.\ \ref{part1:fig19} which thus corresponds to a ``magneto-capillary'' wave turbulence regime. Unlike the above dispersive systems [see Eq.\ (\ref{WKTg}) \& (\ref{WKT}) and \S \ref{spectra}], this $-3$ exponent does not depend on the energy flux exponent $\alpha$. However, the exponent $\alpha$ is still not deduced by dimensional analysis. By measuring the scaling law between the spectrum amplitude $S^{mag}_{\eta}$ and the magnetic field $B$, one then shows that $\alpha=1/3$, {\em i.e.} that the magnetic wave turbulence involves a 4-wave interaction process. One has thus obtain the whole expression of the spectrum of magnetic wave turbulence. 

\begin{figure}[h]
\centerline{
\includegraphics[height=5 cm]{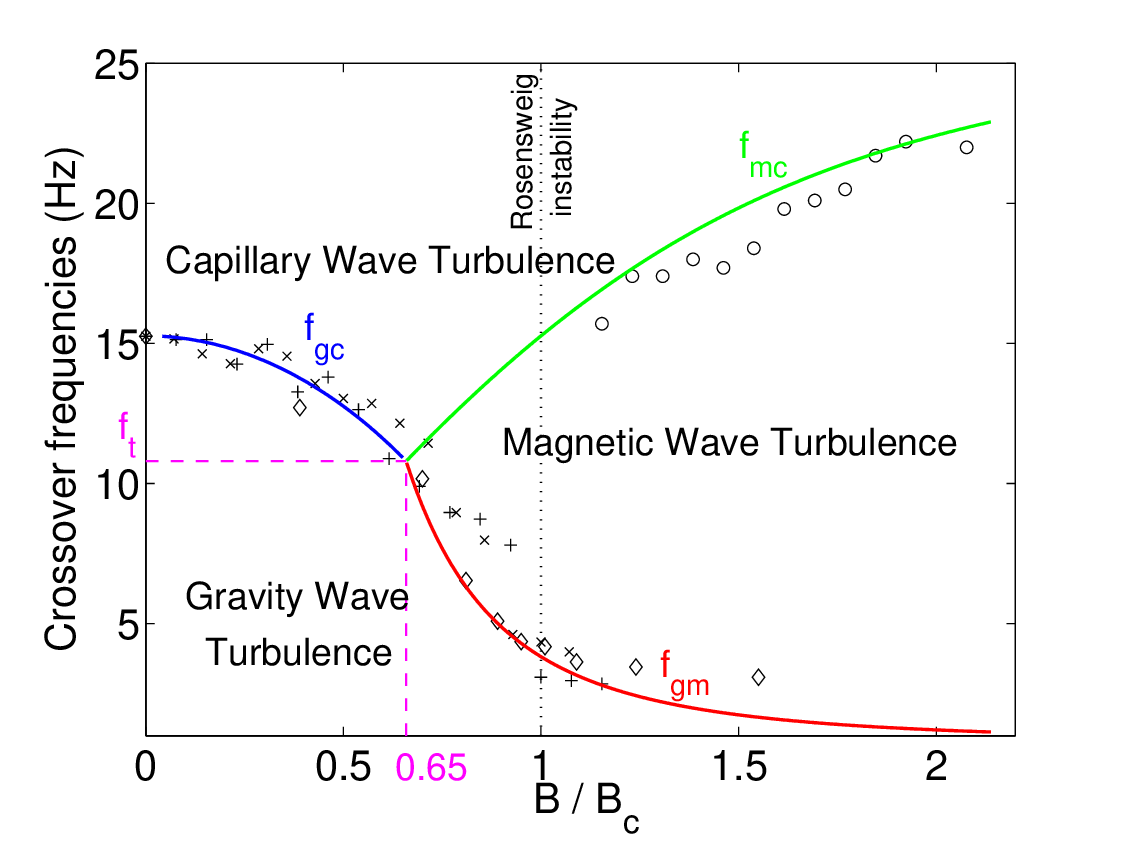}
}
\caption{Crossover frequencies between gravity, magnetic and capillary wave turbulence regimes as a function of the applied magnetic field. Different forcing parameters: ($\times$) 1 to 4 Hz, ($\diamond$) 1 to 5 Hz, and ($+$ or $\circ$) 1 to 6 Hz. Theoretical curves $f_{gc}$, $f_{gm}$ and $f_{mc}$ can be computed analytically from Eq.\ (\ref{rdtheo}) \cite{Boyer08}. The triple point reads $f_t=\frac{1}{2\pi}(\frac{g^3\rho}{\gamma})^{1/4} \simeq 10.8$ Hz and $B_t^2 = \mu_0\sqrt{\rho g \gamma}/f[\chi(H_t)]$, that is $B_t/B_c=0.65$ for our ferrofluid, where $B_c$ is the onset of the Rosensweig instability.}
\label{part1:fig20}
\end{figure}

The existence of the regimes of gravity, magnetic and capillary wave turbulence is reported in the phase space parameters as well as a triple point of coexistence of these three regimes (see Fig.\ \ref{part1:fig20}) \cite{Boyer08}. The theoretical crossover frequency curves, as well as the triple point, can be computed from the dispersion relation of Eq.\ (\ref{rdtheo}), and are found in good agreement with the experimental data with no adjustable parameter (see solid lines in Fig.\ \ref{part1:fig20}). Finally, the statistical distribution of the wave amplitudes and their change with the magnetic field has been also studied: our measurements confirm the nonlinear nature of the wave interaction in a wave turbulence regime, and also show the onset of existence of the magnetic waves \cite{Boyer08}. 

\section{Conclusions and open problems}
\label{objectifs}
I have first presented the context and the state of the art in wave turbulence. I have then focused on the wave turbulence on the surface of a fluid, and I have discussed the main experimental results of our group on this subject. We have measured, for the first time, the key governing parameter of wave turbulence: the energy flux that drives the waves and cascades to small scales through nonlinear interactions. Fluctuations much larger than the mean value have been observed as well as instantaneous negative energy flux events that occurs with a fairly large probability. Taking into account these energy flux fluctuations in theoretical models of cascades remains an open problem.  The scaling of the power spectra of capillary wave turbulence is found in agreement with weak turbulence theory both in laboratory experiments or in low-gravity environment (to study pure capillary waves). A disagreement with weak turbulence theory has been observed for the gravity wave spectrum. It has been also found that gravity waves are intermittent, {\it i.e.} the statistics of velocity increments of the interface are strongly non Gaussian at small scales. This first observation of intermittency in wave turbulence could be related to the large fluctuations of the energy flux. Finally, magnetic wave turbulence has been observed on the surface of a magnetic fluid (ferrofluid) as well as a triple point of coexistence of gravity, capillary and magnetic wave turbulence regimes. These features are understood using dimensional analysis, but no weak turbulence prediction exists for this magnetic case.


 As emphasized in this review, there has been a renewal of interest in wave turbulence since last years. However, open problems are still numerous. It is crucial to consider in a precise way the range of validity of theoretical predictions and to study experimental effects associated to breakdown of existing theories. Finite size effects in experiments lead to a quantization of wave number that can strongly affect energy transfers. Understanding the observed dependence of the frequency-exponent of the gravity wave spectrum with the forcing parameters is thus of primary interest. Besides, most \emph{in situ} or laboratory measurements on wave turbulence are local in space whereas theoretical predictions often concern Fourier space, \emph{e.g.} power spectra or probability density functions of the fluctuations of the Fourier amplitude of a wave component at a given wave number. Thus, an important challenge is to develop a direct probe in Fourier space. Some recent studies measure the amplitudes of gentle interacting waves on a spatial zone  \cite{Moisy,Snouck08,Wright96,Wright97}. However, these optical methods cannot measure amplitudes of steep nonlinear waves as involved in wave turbulence. Nevertheless, one need to access to such steep wave amplitude in the Fourier space in order to get a better understanding of the elementary dynamical processes involved in the energy cascade.  An other challenge is to observe a possible inverse cascade in wave turbulence that has been predicted theoretically \cite{Zakharov67Grav,ZakharovLivre} and numerically \cite{Korotkevitch08}. Finally, a significant issue is to understand the physical mechanisms of intermittency in wave turbulence that has been recently observed \cite{Falcon07b}. 

\section*{Acknowledgements}I thank S. Fauve and C. Laroche who have been strongly involved in the experiments of wave turbulence on the surface of a fluid, as well as S. Auma\^{\i}tre, U. Bortolozzo, F. Boyer and C. Falc\'on \cite{Boyer08,Falcon07,Falcon07b,Falcon08,Falcon08sub2,Falcon08sub1}. This work has been supported by ANR Turbonde BLAN07-3-197846, and by CNES.

\medskip
Received xxxx 20xx; revised xxxx 20xx.
\medskip

\end{document}